# A Summary of COVID-19 Datasets

Syed Raza Bashir[1], Shaina Raza[2], Vidhi Thakkar[3], Usman Naseem[4]

[1]Department of Computer Science, Toronto Metropolitan University, Toronto, Canada
Syedraza.bashir@ryerson.ca
[2]University of Toronto, Canada
Shaina.raza@utoronto.ca
[3]Institute of Aging, Faculty of Nursing, University of Victoria, Victoria, Canada
vidhithakkar@uvic.ca
[4]Department of Information Technology, Sydney International School of Technology and Commerce, Sydney, Australia
usman.n@sistc.nsw.edu.au

## ABSTRACT

*This research presents a review of main datasets that are developed for COVID-19 research. We hope this collection will continue to bring together members of the computing community, biomedical experts, and policymakers in the pursuit of effective COVID-19 treatments and management policies. Many organizations, such as the World Health Organization (WHO) , John Hopkins , National Institute of Health (NIH) , COVID-19 open science table  and such, in the world, have made numerous datasets available to the public. However, these datasets originate from a variety of different sources and initiatives. The purpose of this research is to summarize the open COVID-19 datasets to make them more accessible to the research community for health systems design and analysis. We also discuss the numerous resources introduced to support text mining applications throughout the COVID-19 literature; more precisely, we discuss the corpora, modelling resources, systems, and shared tasks introduced for COVID-19.*

## KEYWORDS

*COVID-19, Text Mining, Public health, Risk, Public Health, COVID-19 Data; Data Science*

## 1. INTRODUCTION

COVID-19, a SARS coronavirus 2-related acute respiratory infectious disease, was discovered in late 2019 and declared a global pandemic by the WHO in March 2020 (4). Recent estimates show that there have been over 500,000 deaths and 400 million jobs lost due to the COVID-19 pandemic (5), causing widespread socioeconomic, political, and health policy implications as well as increases in physical and mental health conditions (6). Many empirical and theoretical studies have looked into COVID-19's morbidity and mortality rates since its discovery (7). The effects of COVID-19 have lasted for over a year since it was first reported. Many people's mental health has been severely impacted by the COVID-19 epidemic. The long-COVID effects are also being studied in the latest research (8). To summarize, the amount of information on the epidemic that would define a generation is becoming nearly overwhelming.

A few projects, such as CoronaNet [1]; CORD-19 [2] and a United States (US) county-level dataset [3], among a few, that presents a collection of data from different sources. However, the nature of these datasets are different, for example CoronaNet [1] and county-level dataset [3] are repositories of government responses, while the CORD-19 [2] is a collection of academic literature. Nonetheless, our approach to this work is unique in that we also focus on multiple data sources and data types, such as epidemiology, literature, government policies, lab's data, mortality and other miscellaneous data. We believe that having a summary of COVID-19 datasets is a necessary first step in the innovation and research landscape for understanding the COVID-19 pandemic and mitigating its effects.

To deal with the overwhelming amount of COVID-19 literature, the computer community has developed text mining corpora, modelling resources, systems, and community-wide shared tasks. Text mining uses corpora, which are collections of documents pre-processed to extract machine-readable text. Corpora are collections of documents that have been pre-processed to extract machine-readable text and are used for text mining; in this example, the corpora contain scientific publications. Text mining practitioners can include modelling resources into production systems. These resources include text embeddings, data annotations, pretrained language models, and knowledge graphs.

In this article, we discuss some of the most common types of COVID-19 data sources available. We also outline prominent text mining systems that have been developed/implemented to assist text mining in the COVID-19 literature. Our goal is to make it easier for researchers, scientists, and other interested parties to get most up-to-date data for their research. We believe that a list of the most relevant sources of COVID-19 data is needed for the purpose of health systems design and health services research. It is our hope that this resource will continue to bring together the computing community, biomedical experts, and policy makers in the pursuit of effective treatments and management policies for Covid-19.

## 2. COVID-19 DATASETS

We categorize the types of COVD-19 related into different types as per the type of research being conducted. Each of this dataset below is supported by a reference (a reference for dataset, a link to the dataset page and the main provider/source of dataset).

### 2.1. Survey Data

Survey data is defined as the outcome data collected from a sample of survey respondents [4]. This data is extensive information gathered from a specific target audience about a specific topic to conduct research. We present some of the survey data related to COVID-19, in Table 1, to help researchers better understand the influences on participants' daily changes, notably the impact of COVID-19 impacts.

Table 1: Survey data (each underline data source is hyperlink)

| Data | Source |
|---|---|
| - Canadian Perspectives Survey Series (CPSS) [5]<br>- Collection time: January 15, 2020 to March 15, 2021<br>- The CPSS entails assembling a group of people who agree to take a series of brief online surveys related to COVID-19 over the course of a year.<br>- It consists of following surveys<br>- COVID-19 in Canada: An update on social and economic impacts, Fall 2021 [6]<br>- Impacts of COVID-19 [7]<br>- COVID-19 and working from home, 2020 [8]<br>- Impacts of COVID-19 on job security and personal finances, 2020 [9]<br>- Canadians report lower self-perceived mental health during the COVID-19 pandemic [10]<br>- Mental health of Canadians during the COVID-19 pandemic [11]<br>- Canadian Perspectives Survey Series 2: Monitoring the effects of COVID-19, May 2020 [12]<br>- Food insecurity before and during the COVID-19 pandemic [13] | Statistic Canada [19] |
| - Food insecurity during the COVID-19 pandemic, May 2020 [13]<br>- Canadian Perspectives Survey Series 3: Resuming economic and social activities during COVID-19 [14]<br>- Precautions that Canadians will take or continue to take as COVID-19 safety measures are relaxed [15]<br>- Canadian Perspectives Survey Series 4: Information sources consulted during the pandemic, July 2020 [16]<br>- Canadians spend more money and time online during pandemic and over two-fifths report a cyber incident [17]<br>- COVID-19 in Canada: A Six-month Update on Social and Economic Impacts [18] | |
| Canadian COVID-19 Antibody and Health Survey (CCAHS)<br>Collection time: From November 2, 2020, to April 16, 2021<br>This survey gathers data in two parts. The first section is an automated questionnaire concerning general health and COVID-19 exposure. The second component is an at-home finger-prick blood test, which is sent to a lab to determine the existence of COVID-19 antibodies. | Statistic Canada |
| Pulse Survey on COVID-19 and its Impacts on Statistics Canada Employees (PSCISCE) [20]<br>Collection time: April 27, 2021<br>The goal of this poll is to analyze the workforce's health in near real time, give immediate information on the COVID-19 crisis's impact, and shape the development of tools and action plans to better support our employees going ahead. | Statistic Canada |
| Uniform Crime Reporting Survey – Selected police reported crime statistics – Special COVID-19 report to Statistics Canada [21]<br>Collection time: January 15 to December 31, 2022<br>The purpose of this survey is to provide timely monthly aggregate data on the types of police-reported activities, including criminal occurrences and other requests for police assistance, that happened during the COVID-19 pandemic's initial months. The data can be used by policymakers, researchers, and the general public. | Statistic Canada |

## 2.2. Epidemiology Data

Epidemiology is the study of how often diseases occur in different groups of people and why they occur [22]. When developing and evaluating disease prevention strategies, epidemiological information is used as a guide for the management of patients who have already developed the disease. Similarly, to the clinical findings and pathology of a disease, the epidemiology of a disease is an integral part of the disease's fundamental description. The subject has its unique techniques for data collection and interpretation, that must be understood.

Table 2: Epidemiology data

| Data | Source |
|---|---|
| COVID-19 daily epidemiology update [23]<br>This data consists of COVID-19 case summaries from across Canada and over time. it also presents a summary of hospitalizations and deaths, testing, variants of concern, and exposures is provided. | Canada.ca [24], Canada COVID-19 hub [25] |

| WHO daily data [26] This dashboard presents official daily counts of COVID-19 cases and deaths worldwide. | World Health Organization (WHO) [27] |
|---|---|
| COVID-19 Canada Open Data Working Group [28], [29] This is an epidemiological data from the COVID-19 Epidemic in Canada. | Open COVID [30] |
| Covid-19 Cases in Toronto [31] This dashboard gives information about case counts, epidemiological summary of cases, active outbreaks and vaccine data in Toronto. | City of Toronto [32] |

## 2.3. COVID-19 Dataset by Source

We also gathered some data sources that provide COVID-19 statistics by country, province and city. The goal is that one may use the original data sources from the government websites, rather than relying on different other sources. In the rest of this review article, we present the main source of data, more details about each dataset can be found in the respective paper or source.

Table 3: Data sources by country and states.

| **Data** | **Source** |
|---|---|
| COVID-19 cases and forecasting hot spots, country wise [33] | Mayo Clinic [34] |
| COVID Data Tracker [35] | Centers for Disease Control and Prevention [36] |
| NYTimes Latest Map and Case Count [37] | New York Times [38] |
| COVID-19 Cases by Country [39] | Johns Hopkins [40] |
| COVID Risk & Vaccine Tracker [41] | COVID ActNow [41] |
| Global COVID-19 Tracker [42] | Kaiser Family Foundation [43] |
| COVID Tracking [44] | COVID Tracking Project [45] |
| ACLED Data Export Tool [46] | Armed Conflict Location & Event Data Project [47] |
| COVID-19 cases daily [48] | European Centre for Disease Prevention and Control [48] |
| 60 million anonymized COVID-19 case [49] | Global Health [49] |

## 2.4. COVID-19 Projections Data

The COVID-19 projection models work with a big dataset. When there are larger datasets, there exists the possibility to study statistical changes in the epidemiology, pathogenesis, and spread of the COVID-19 pandemic within a country and across provinces. This type of information is valuable to policymakers in creating predictive models and in order to inform public policy restrictions on wearing masks, maintaining physical distance, and advising on how the economy should function, including rules around the conduct of small businesses and restaurants.

COVID-19 projects data can inform policy changes, help to measure the spread of the pandemic amongst population groups and inform the calculation of mortality rates. Although historical data is rarely a perfect predictor of the future, when used correctly, it can provide some insight into what the coming days and weeks may look like. It is also possible to do predictive modelling with Tableau and SPSS software that can help predictive logarithmic models to forecast the anticipated increase or decrease in cases over a time period.

Table 4: Data sources by country and states.

| Data | Source |
|---|---|
| COVID-19 projections bi-weekly [50] | Institute for Health Metrics and Evaluation [51] |
| COVID-19 Staffing Simulator [52] | LEVRUM [53] |
| State-level social distancing policies [54] | University of Washington [55] |
| Alcohol Sales During the COVID-19 Pandemic [56] | Alcohol Policy for Information Systems [57] |
| COVID-19 prison data [58] | Marshall Project [59] |

### 2.5. Lab Data

The number of COVID-19 positive tests tell us only so much about how the pandemic is behaving in each community. This is due, in large part, to testing bias, in which tests are limited to the sickest and most stereotypically presenting patients. To fully comprehend how deeply embedded COVID-19 is in a given community, the total volume of tests must be reported alongside the ever-present positive case count.

Table 5: COVID-19 Lab Datasets

| Data | Source |
|---|---|
| Virology COVID-19 Dashboard [60] | University of Washington Medicine's Department of Laboratory Medicine [55] |
| Online access to COVID-19 Vaccination Information and COVID-19 lab test results [61] | eHealth Ontario [62] |
| Vaccine Adverse Event Reporting System [63] | Vaccine Adverse Event Reporting System [64] |
| Public Health Ontario Data and Analysis [65] | Public Health Ontario [66] |
| BCCDC COVID-19 [67] | British Columbia Centre for Disease Control [68] |

### 2.6. Mobility Data

Mobility data can also be used to determine the extent to which communities have been shut down. There are six types of areas that Google looks at in its dashboard, which may be broken down by country or region. These include retail and recreation; grocers and pharmacy; parks; transit stations; businesses; and residential sites.

Table 6: COVID-19 Travel and Mobility Data

| Data | Source |
|---|---|
| COVID-19 Mobility data [69] | Google [70] |
| COVID-19 Reports [71] | Institute for Disease Modeling [72] |
| Facebook Data for Good [73] | Meta [74] |

### 2.7. Mortality Data

COVID-19 mortality is defined as those people who died because of complications from COVID-19 as well as with other ailments because of their incapacity or reluctance to access a possibly overwhelmed health system, or because they did not have the right treatment in the system at the time, is also an important data. Plaxovid and retonovir, for example, were unavailable for a long time and resulted in numerous deaths.

Table 7: Mortality Data

| Data | Source |
|---|---|
| US COVID-19 vs other causes of deaths [75] | Flourish [76] |
| Canada COVID-19 mortality data [77] | Worldometers [78] |

## 2.8. Public and Mental Health Datasets

We also gathered public health and mental health COVID-19 data from miscellaneous sources, which are given below.

Table 8: Miscellaneous sources of Data

| Data | Source |
|---|---|
| Michigan COVID-19 Data [79] | Michigan Government [80] |
| COVID-19 Public datasets [81] | Google [70] |
| Unintended consequences of COVID-19: Impact on harms caused by substance use [82] | Canadian Institute for Health Information |
| Unintended consequences of COVID-19: Impact on self-harm behavior [83] | Canadian Institute for Health Information |
| Ontario Mental Health Reporting System Metadata [84] | Canadian Institute for Health Information |
| Mental Health During COVID-19 Outbreak [85] | Centre for Addiction and Mental Health [86] |
| Impact of COVID-19 on accidental falls [87] | Canadian Institute for Health Information |
| Flatten: COVID-19 Survey Data on Symptoms, Demographics and Mental Health in Canada [88] | PhysioNet [89] |
| COVID-19 National Survey [90] | Centre for Addiction and Mental Health [86] |
| Provincial COVID-19 Vaccine (COVaxON) [91] | Ontario Health Data Platform [92] |
| COMPASS CIHR mental health [93] | COMPASS SYSTEM [94] |

## 3. TEXT MINING DATASETS AND TOOLS

Text mining, also known as text data mining or text analytics, is a method for extracting valuable information from text [95]. As a result of the COVID-19 pandemic, several datasets containing full text about COVID-19, SARS-CoV-2, and related coronaviruses have been released. These freely available datasets are being made available to the global research community to enable the application of recent advances in natural language processing and other artificial intelligence (AI) techniques in order to generate new insights to aid in the ongoing fight against this infectious disease.

### 3.1. Text Mining Datasets

The COVID-19 Open Research Dataset [2] is one of the earliest and largest literature corpora created to support COVID-19 text mining. It is a corpus of metadata and full text of COVID-19 publications and preprints released daily by Semantic Scholar at the Allen Institute for AI in collaboration with Microsoft Research, IBM Research, Kaggle, the Chan-Zuckerberg Initiative, the National. This corpus was first made available on March 16, 2020, at the request of the White House Office of Science and Technology Policy, in order to support community-wide efforts to apply text mining techniques to coronavirus literature.

The corpus includes papers from PubMed Central (PMC), PubMed, the World Health Organization's COVID-19 database and preprint servers bioRxiv, medRxiv, and arXiv. Paper metadata from these sources is synchronised, PDFs are converted to machine-readable JSON using the S2ORC pipeline described in [96]. As of September 15, 2020, the corpus contained over 260 000 paper entries (with 105 000 full text entries).

LitCovid is a curated set of open access COVID-19 papers from PubMed, currently containing more than 240,000 papers and growing. LitCovid is focused on tracking publications specific to COVID-19, while CORD-19 captures the coronavirus literature more broadly, including other coronaviruses (e.g. SARS and MERS) and a wider time period (i.e. before the current outbreak). LitCovid does NOT include pre-prints. LitCovid only includes relevant articles from PubMed.

The other well-known COVID-19 literature repositories are World Health Organization (WHO) COVID-19 database [97] and the Centers for Disease Control and Prevention's COVID-19 research articles database [98]. These databases overlap with other corpora; for example, the WHO database is ingested by CORD-19, and much of the CDC database overlaps with PubMed and PMC, both of which are sources of papers in CORD-19 and LitCovid. The CDC database also includes a collection of white papers and technical reports. Finally, several publishers have compiled and released collections of their COVID-19 literature, such as Elsevier's Novel Coronavirus Information Center [99], Springer Nature's Coronavirus Research Highlights [100], and other publishers who provide literature under temporary open access licenses. We summarize these text mining datasets related to COVID-19 in Table 9.
Text

Table 9: Text Mining Data

| Data | Source |
| --- | --- |
| CORD-19: COVID-19 Open Research Dataset [2] | Allen Institute for AI [101] |
| Research COVID-19 with AVOBMAT [102] | AVOBMAT [103] |
| LitCovid [104] | NLM/NCBI BioNLP Research Group [105] |
| World Health Organization (WHO) COVID-19 database [97] | WHO [27] |
| Centers for Disease Control and Prevention's COVID-19 research articles database [98] | CDC [36] |
| Elsevier's Novel Coronavirus Information Center [99], | Elsevier Publisher |
| Springer Nature's Coronavirus Research Highlights [100] | Springer Nature Publisher |

## 3.2. Text Mining Tools

Many text mining systems for COVID-19 literature have been released. We compile a list of important tools in Table 10.

Table 10: Text Mining Tools

| Tool | Data source |
| --- | --- |
| Covidex [106] | CORD-19, ClinicalTrials.gov |
| KDCovid [107] | CORD-19 |
| Azure Cognitive Search [108] | CORD-19 |
| CADTH [109] | Multiple sources |

## 4. CONCLUSIONS

Since the COVID-19 emergence, research institutes and governments have made numerous databases publicly available to allow research groups (and independent individuals) to analyse data relating to the COVID-19's spread [110]. These databases are dispersed across a variety of initiatives and sources. The purpose of this article is to compile a list of all of the world's major open databases and data initiatives. Our goal is to provide a road map for establishing links between various scientific fields (health science, epidemiology, artificial intelligence, and mental health). We provide most up-to-date information for a wide range of COVID-19 datasets, which are critical for knowledge synthesis in evidence-based medicine. This review of COVID-19 research can benefit researchers, healthcare professionals, and the general public. Each of these sources is open source and can be downloaded and used in a variety of computational tools and systems.

While this paper includes a list of the most commonly used COVID-19 datasets and tools, it may not include many other COVID-19 datasets. However, because we mainly curated the most important and up-to-date COVID-19 data sources, we believe this list is representative of other COVID-19 datasets as well. We put forward this research direction and give researchers the opportunity to look for more data sources and collaborate to fight the pandemic.

## ACKNOWLEDGEMENTS

We would like to acknowledge that this research and manuscript is a part of my CIHR Health Systems Impact Fellowship.

## REFERENCES

[1] C. Cheng, J. Barceló, A. S. Hartnett, R. Kubinec, and L. Messerschmidt, "COVID-19 Government Response Event Dataset (CoronaNet v.1.0)," *Nat. Hum. Behav.*, vol. 4, no. 7, pp. 756–768, 2020.

[2] L. Lu Wang et al., "CORD-19: The Covid-19 Open Research Dataset.," 2020.

[3] B. D. Killeen et al., "A County-level Dataset for Informing the United States' Response to COVID-19," 2020.

[4] S. G. Heeringa, B. T. West, and P. A. Berglund, *Applied survey data analysis*. chapman and hall/CRC, 2017.

[5] Statistics Canada, "Canadian Perspectives Survey Series, 2020," 2020. [Online]. Available: https://www.statcan.gc.ca/en/survey/household/5311. [Accessed: 23-Jan-2022].

[6] Statistics Canada, "COVID-19 in Canada: A One-year Update on Social and Economic Impacts," no. March, pp. 1–52, 2021.

[7] Statistics Canada, "Canadian Perspectives Survey Series 1 : Impacts of COVID-19," *Stat. Canada*, pp. 9–14, 2020.

[8] T. Daily, "Working from home during the COVID-19 pandemic Working from home: a new experiment for many Canadian workers and employers," no. April 2020, 2021.

[9] Statistics Canada, "Canadian Perspectives Survey Series 1: personal finances, 2020," *Dly.*, vol. 11-001–X, pp. 17–20, 2020.

[10] H. Gilmour, "COVID-19: Data to Insights for a Better Canada Canadians report lower self-perceived mental health during the COVID-19 pandemic," *Stat. Canada*, no. 45280001, 2020.

[11] S. Canada, "Statistics:Covid :Mental Health," no. May, p. 2020, 2020.

[12] Statistics Canada, "Canadian Perspectives Survey Series 2: Monitoring the effects of COVID-19," *The Daily*, 2020.

[13] Statistics Canada, "Food insecurity before and during the COVID-19 pandemic , 2017 / 2018 and May 2020," no. May, p. 8300, 2020.

[14] Statistics Canada, "Canadian Perspectives Survey Series 3: Resuming economic and social activities during COVID-19," *Dly.*, 2020.

[15] S. Canada, "PRECAUTIONS THAT CANADIANS WILL TAKE OR CONTINUE TO TAKE AS COVID-19 SAFETY MEASURES ARE RELAXED," p. 35446, 2020.

[16] Statistics Canada, "Canadian Perspectives Survey Series 4: pandemic, July 2020," *Dly.*, vol. 11-001–X, no. July, pp. 17–20, 2020.

[17] T. Daily, "Canadians spend more money and time online during pandemic and over two-fifths report a cyber incident," pp. 40–43, 2020.

[18] A. Arora, "COVID-19 in Canada: A Six-month Update on Social and Economic Impacts," no. September, pp. 1–36, 2020.

[19] Government of Canada, "Statistics Canada: Canada's national statistical agency," 2016. [Online]. Available: http://www.statcan.gc.ca/pub/62f0026m/2016001/chap4-eng.htm. [Accessed: 23-Jan-2022].

[20] Statistics Canada, "Pulse Survey on COVID-19 and its Impacts on Statistics Canada Employees (PSCISCE)." [Online]. Available: https://www.statcan.gc.ca/en/survey/household/5326. [Accessed: 23-Jan-2022].

[21] Statistics Canada, "Uniform Crime Reporting Survey - Selected police reported crime statistics - Special COVID-19 report to Statistics Canada." [Online]. Available: https://www.statcan.gc.ca/en/survey/business/3302. [Accessed: 23-Jan-2022].

[22] The British Medical Journal, "Epidemiology for the uninitiated," *BMJ : British Medical Journal*, 2018. [Online]. Available: https://www.bmj.com/about-bmj/resources-readers/publications/epidemiology-uninitiated/6-ecological-studies?hwshib2=authn%3A1535394580%3A20180826%253Ae48fec14-6539-4bab-9bd2-b60a7372a9d1%3A0%3A0%3A0%3ANhpf%2BmFJqPMhlDoGoRukJw%3D%3D. [Accessed: 19-Jan-2022].

[23] Public Health Agency of Canada, "COVID-19 Daily Epidemiology Update," *Coronavirus disease (COVID-19): For health professionals*, 2021. [Online]. Available: https://www.canada.ca/content/dam/phac-aspc/documents/services/diseases/2019-novel-coronavirus-infection/surv-covid19-epi-update-eng.pdf.

[24] Canada.ca, "Home - Canada.ca," 2021. [Online]. Available: https://www.canada.ca/en.html. [Accessed: 23-Jan-2022].

[25] Public Health Agency of Canada, "COVID-19 Daily Epidemiology Update," *Coronavirus disease (COVID-19): For health professionals*, 2021. [Online]. Available: https://www.canada.ca/content/dam/phac-aspc/documents/services/diseases/2019-novel-coronavirus-infection/surv-covid19-epi-update-eng.pdf. [Accessed: 23-Jan-2022].

[26] E. Dong, H. Du, and L. Gardner, "An interactive web-based dashboard to track COVID-19 in real time," *Lancet Infect. Dis.*, vol. 20, no. 5, pp. 533–534, 2020.

[27] A. EGDAHL, "WHO: World Health Organization.," *The Illinois medical journal*, 1954. [Online]. Available: https://www.who.int/. [Accessed: 23-Jan-2022].

[28] I. Berry, J. P. R. Soucy, A. Tuite, and D. Fisman, "Open access epidemiologic data and an interactive dashboard to monitor the COVID-19 outbreak in Canad," *CMAJ*, vol. 192, no. 15, p. E420, Apr. 2020.

[29] I. Berry *et al.*, "A sub-national real-time epidemiological and vaccination database for the COVID-19 pandemic in Canada," *Sci. Data*, vol. 8, no. 1, Dec. 2021.

[30] Opencovid, "COVID-19 Canada Open Data Working Group." [Online]. Available: https://opencovid.ca/. [Accessed: 23-Jan-2022].

[31] T. Publich Health, "COVID-19 Cases in Toronto - City of Toronto Open Data Portal," 2020. [Online]. Available: https://open.toronto.ca/dataset/covid-19-cases-in-toronto/. [Accessed: 23-Jan-2022].

[32] N. Services, "City of Toronto," *Policy Analysis*, 2004. [Online]. Available: https://www.toronto.ca/. [Accessed: 23-Jan-2022].

[33] Mayo Clinic, "U.S. COVID-19 Map: Tracking the Trends." [Online]. Available: https://www.mayoclinic.org/coronavirus-covid-19/map. [Accessed: 27-Jan-2022].

[34] Mayo Clinic, "Mayo Clinic," *Mayo Clinic*, 2019. [Online]. Available: https://www.mayoclinic.org/diseases-conditions/hepatocellular-carcinoma/cdc-20354552. [Accessed: 27-Jan-2022].

[35] CDC, "CDC COVID Data Tracker," *Centers for Disease Control and Prevention*, 2020. [Online]. Available: https://covid.cdc.gov/covid-data-tracker/#vaccinations_vacc-total-admin-rate-total%0Ahttps://covid.cdc.gov/covid-data-tracker/#vaccinations-pregnant-

women%0Ahttps://covid.cdc.gov/covid-data-tracker/#datatracker-home%0Ahttps://covid.cdc.gov/covid-data-trac. [Accessed: 27-Jan-2022].

[36] C. for D. C. and Prevention., "Centers for disease control and prevention," *Indian Journal of Pharmacology*, 2004. [Online]. Available: https://www.cdc.gov/. [Accessed: 27-Jan-2022].

[37] T. N. Y. Times, "Covid in the U.S.: Latest Map and Case Count," *The New York Times*, 2020. [Online]. Available: https://www.nytimes.com/interactive/2021/us/covid-cases.html. [Accessed: 27-Jan-2022].

[38] A. Cowell, "The New York Times - Breaking News, US News, World News and Videos," *The New York Times*, 2003. [Online]. Available: https://www.nytimes.com/. [Accessed: 27-Jan-2022].

[39] Johns Hopkins University, " COVID-19 United States Cases by County Johns Hopkins University," *Https://Coronavirus.Jhu.Edu/Us-Map*, 2021. [Online]. Available: https://coronavirus.jhu.edu/us-map. [Accessed: 27-Jan-2022].

[40] Johns Hopkins University, "Mortality Analyses - Johns Hopkins Coronavirus Resource Center," *Johns Hopkins Coronavirus Resource Center*, 2020. [Online]. Available: https://coronavirus.jhu.edu/data/mortality. [Accessed: 27-Jan-2022].

[41] C. A. Now, "Realtime U.S. COVID Map & Vaccine Tracker - Covid Act Now." [Online]. Available: https://covidactnow.org/?s=22401166. [Accessed: 27-Jan-2022].

[42] G. C.-19 Tracker, "Global COVID-19 Tracker – Updated as of January 27 | KFF." [Online]. Available: https://www.kff.org/coronavirus-covid-19/issue-brief/global-covid-19-tracker/. [Accessed: 27-Jan-2022].

[43] Kaiser Family Foundation, "KFF - Health Policy Analysis, Polling, Journalism and Social Impact Media." [Online]. Available: https://www.kff.org/. [Accessed: 27-Jan-2022].

[44] CovidTracking.com, "The COVID Tracking Project | The COVID Tracking Project," *The Atlantic Monthly Group*, 2020. [Online]. Available: https://covidtracking.com/. [Accessed: 27-Jan-2022].

[45] The Atlantic, "The Covid Tracking Project API," *The Covid Tracking Project*, 2020. [Online]. Available: https://covidtracking.com/api. [Accessed: 27-Jan-2022].

[46] ACLED, "Data Export Tool | ACLED," 2021. [Online]. Available: https://acleddata.com/data-export-tool/.

[47] ACLED, "Home - ACLED," 2021. [Online]. Available: https://acleddata.com/#/dashboard.

[48] R. Evans, "European Centre for Disease Prevention and Control," *Nursing standard (Royal College of Nursing (Great Britain) : 1987)*, 2014. [Online]. Available: https://www.ecdc.europa.eu/en. [Accessed: 27-Jan-2022].

[49] Global.health, "Global.health: A Data Science Initiative - Global.health," 2021.

[50] IHME, "COVID-19." 2022.

[51] M. Arthur, "Institute for Health Metrics and Evaluation," *Nursing Standard*, vol. 28, no. 42. pp. 32–32, 2014.

[52] LEVRUM, "Levrum COVID-19 Simulator," 2021. [Online]. Available: https://covidsim.levrum.com/.

[53] LEVRUM, "Home - Levrum Data Technologies." [Online]. Available: https://levrum.com/.

[54] C. Wu, F. Wu, Y. Chen, S. Wu, Z. Yuan, and Y. Huang, "Neural Metaphor Detecting with CNN-LSTM Model," 2018, pp. 110–114.

[55] University of Washington, "UW Homepage." 2016.

[56] APIS, "About Alcohol Policy | APIS - Alcohol Policy Information System." 2021.

[57] National Institute on Alcohol Abuse and Alcoholism, "Alcohol Policy Information System (APIS)," vol. 2008, no. December 19. 2005.

[58] K. Park and T. Meagher, "A State-By-State Look at 15 Months of Coronavirus in Prisons | The Marshall Project," *The Marshall Project*. 2021.

[59] Marshall Project, "The Marshall Project." 2021.

[60] U. Medicine, "UW Virology COVID-19 Dashboard." 2020.

[61] eHealth Ontario-Ministry of Health, "Online access to COVID-19 lab test results for Health Care Providers | eHealth Ontario | It's Working For You." .

[62] eHealth Ontario, "eHealth Ontario | It's Working For You." 2017.

[63] "VAERS - Data," 2021. [Online]. Available: https://vaers.hhs.gov/data.html.

[64] VAERS, "The Vaccine Adverse Event Reporting System (VAERS)," *Vaccine*, vol. 12, no. 10. p. 960, 1994.

[65] Public Health Ontario, "Coronavirus Disease 2019 (COVID-19) – PCR | Public Health Ontario," *Public Health Ontario*. 2020.

[66] K. Wong and A. Piatkowski, "Public Health Ontario." pp. 1–30, 2016.
[67] BCCDC, "COVID-19 – BCCDC Foundation for Public Health." 2022.
[68] Infections, "BC Centre for Disease Control." 2009.
[69] Google, "COVID-19 Community Mobility Reports." 2022.
[70] Google, "Google," 2022. [Online]. Available: https://www.google.com/.
[71] IDMOD COVID-19, "COVID Reports | IDMOD." 2022.
[72] Institute of Disease Modeling, "Home Page | IDMOD." 2022.
[73] J. M. Wing, "Data for Good," 2018. [Online]. Available: https://dataforgood.facebook.com/.
[74] Meta, "Welcome to Meta | Meta," *Meta*. p. 13, 2021.
[75] Flourish, "Covid vs. US Daily Average Cause of Death | Flourish." 2022.
[76] Flourish, "Flourish | Data Visualisation \& Storytelling." 2021.
[77] Worldometer, "Canada COVID - Coronavirus Statistics - Worldometer." .
[78] Worldometers, "Worldometers: Real Time World Statistics," *Choice Reviews Online*, 2012. [Online]. Available: https://www.worldometers.info/.
[79] State of Michigan, "Coronavirus: Michigan Data," *Michigan.gov*. 2020.
[80] "State of Michigan." 2022.
[81] Google, "Google Cloud Platform," *Https://Cloud.Google.Com*. pp. 11–12, 2016.
[82] Canadian Institute for Health Information, "Unintended consequences of COVID-19: Impact on harms caused by substance use." pp. 1–20, 2021.
[83] "Unintended Consequences of COVID-19: Impact on Harms Caused by Substance Use," 2021.
[84] Canadian Institute for Health Information, "Ontario Mental Health Reporting System," *Data Quality Documentation*. 2016.
[85] CAMH, "Mental Health and the COVID-19 Pandemic | CAMH." 2021.
[86] J. C. Negrete, J. Collins, N. E. Turner, and W. Skinner, "The Centre for Addiction and Mental Health," vol. 49, no. 12. 2004.
[87] CIHI, "Impact of COVID-19 on accidental falls in Canada | CIHI." 2021.
[88] Flatten, "Flatten: COVID-19 Survey Data on Symptoms, Demographics and Mental Health in Canada v1.0." 2021.
[89] PhysioNet, "PhysioNet," 2021. [Online]. Available: https://www.physionet.org/.
[90] The Centre for Addiction and Mental Health, "COVID-19 National Survey Dashboard." 2020.
[91] P. D. Yadav *et al.*, "Neutralization of Beta and Delta variant with sera of COVID-19 recovered cases and vaccinees of inactivated COVID-19 vaccine BBV152/Covaxin," *J. Travel Med.*, vol. 28, no. 7, p. taab104, 2021.
[92] OHDP, "OHDP - Researchers - Datasets - OHDP," 2021. [Online]. Available: https://ohdp.ca/datasets/.
[93] M. C. Buchan, I. Romano, A. Butler, R. E. Laxer, K. A. Patte, and S. T. Leatherdale, "Bi-directional relationships between physical activity and mental health among a large sample of Canadian youth: a sex-stratified analysis of students in the COMPASS study," *Int. J. Behav. Nutr. Phys. Act.*, vol. 18, no. 1, pp. 1–11, 2021.
[94] COMPASS, "COMPASS CIHR mental health | Compass System | University of Waterloo," 2021. [Online]. Available: https://uwaterloo.ca/compass-system/compass-system-projects/compass-cihr-mental-health.
[95] A.-H. Tan and others, "Text mining: The state of the art and the challenges," in *Proceedings of the pakdd 1999 workshop on knowledge disocovery from advanced databases*, 1999, vol. 8, pp. 65–70.
[96] K. Lo, L. L. Wang, M. Neumann, R. Kinney, and D. Weld, "{S}2{ORC}: The Semantic Scholar Open Research Corpus," in *Proceedings of the 58th Annual Meeting of the Association for Computational Linguistics*, 2020, pp. 4969–4983.
[97] WHO, "Global research on coronavirus disease (COVID-19) WHO Databse," 2021. [Online]. Available: https://www.who.int/emergencies/diseases/novel-coronavirus-2019/global-research-on-novel-coronavirus-2019-ncov. [Accessed: 30-Dec-2021].
[98] M. Wilson and P. J. K. Wilson, "Coronavirus Disease 2019 (COVID-19)," *Close Encounters of the Microbial Kind*. pp. 185–196, 2021.
[99] W. E. el. E. M. Salazar, J. Barochiner, "Novel Coronavirus Information Center," *Elsevier*, vol. 20. pp. 2–5, 2020.
[100] Springer Nature, "Coronavirus (COVID-19) Research Highlights | For Researchers | Springer Nature." 2022.
[101] allenai.org, "Allen Institute for AI." 2020.

[102] AVOBMAT, "Research COVID-19 with AVOBMAT – AVOBMAT." 2020.
[103] R. Péter, Z. Szántó, J. Seres, V. Bilicki, and G. Berend, "AVOBMAT: a digital toolkit for analysing and visualizing bibliographic metadata and texts." pp. 43–55, 2020.
[104] Q. Chen, A. Allot, and Z. Lu, "LitCovid: An open database of COVID-19 literature," *Nucleic Acids Res.*, vol. 49, no. D1, pp. D1534–D1540, 2021.
[105] Z. Lu, "Zhiyong Lu - NCBI - NLM," 2021. [Online]. Available: https://www.ncbi.nlm.nih.gov/research/bionlp/.
[106] E. Zhang *et al.*, "Covidex: Neural ranking models and keyword search infrastructure for the covid-19 open research dataset," *arXiv Prepr. arXiv2007.07846*, 2020.
[107] KDCovid, "KDCovid." 2020.
[108] Azue, "Azure Cognitive Search - Covid-19 Search Demo." 2020.
[109] CADTH, "CADTH COVID-19 Search Strings - CADTH Covid-19 Evidence Portal." 2021.
[110] L. L. Wang and K. Lo, "Text mining approaches for dealing with the rapidly expanding literature on COVID-19," *Brief. Bioinform.*, vol. 22, no. 2, pp. 781–799, 2021.